\documentclass{natureprintstyle_comment}
\usepackage{amssymb}
\usepackage{graphicx}
\usepackage[none]{hyphenat}
\usepackage{subfigure}
\usepackage[hyphens]{url}
\usepackage{fancyhdr}

\usepackage{scalerel}
\usepackage{tikz}
\usetikzlibrary{svg.path}

\usepackage[style=nature, backend=biber, articletitle=true, date=year]{biblatex}
\newcommand{\msun}{M_{\odot}}
\newcommand{\myr}{M_{\odot}\,{\rm yr^{-1}}}

\newcommand{\mbh}{M_\bullet}
\newcommand{\mbhn}{M_{\bullet,10}}
\newcommand{\txk}{T_{\rm x,0.4}}

\newcommand{\rs}{r_{\rm S}}
\newcommand{\rv}{r_{\rm vir}}

\newcommand{\Chandra}{\textit{Chandra}}
\newcommand{\Athena}{\textit{Athena}}

\newcommand{\Newton}{\textit{XMM-Newton}}
\newcommand{\Hitomi}{\textit{Hitomi}}

\newcommand{\Magellan}{\textit{Magellan}}
\newcommand{\Suzaku}{\textit{Suzaku}}
\newcommand{\NuSTAR}{\textit{NuSTAR}}
\newcommand{\JWST}{\textit{JWST}}
\newcommand{\XRISM}{\textit{XRISM}}
\newcommand{\ELT}{\textit{ELT}}
\newcommand{\SOFIA}{\textit{SOFIA}}
\newcommand{\SKA}{\textit{SKA}}
\newcommand{\MUSE}{\textit{MUSE}}
\newcommand{\ALMA}{\textit{ALMA}}
\newcommand{\HST}{\textit{HST}}
\newcommand{\EHT}{\textit{EHT}}
\newcommand{\LSST}{\textit{LSST}}

\newcommand{\LISA}{\textit{LISA}}

\def\stacksymbols #1#2#3#4{\def\theguybelow{#2}
        \def\verticalposition{\lower#3pt}
        \def\spacingwithinsymbol{\baselineskip0pt\lineskip#4pt}
        \mathrel{\mathpalette\intermediary#1}}
\def\intermediary #1#2{\verticalposition\vbox{\spacingwithinsymbol
        \everycr={}\tabskip0pt
        \halign{$\mathsurround0pt#1\hfil##\hfil$\crcr#2\crcr
                \theguybelow\crcr}}}

\def\kms{{\rm km\,s^{-1}}}

\def\R500{$R_{500}$}
\def\r500{R_{500}}

\fancypagestyle{plain}{
\fancyhf{}
\fancyfoot[R]{Nature Astronomy Comment $|$~~\thepage}

}
\definecolor{orcidlogocol}{HTML}{A6CE39}
\tikzset{
  orcidlogo/.pic={
    \fill[orcidlogocol] svg{M256,128c0,70.7-57.3,128-128,128C57.3,256,0,198.7,0,128C0,57.3,57.3,0,128,0C198.7,0,256,57.3,256,128z};
    \fill[white] svg{M86.3,186.2H70.9V79.1h15.4v48.4V186.2z}
                 svg{M108.9,79.1h41.6c39.6,0,57,28.3,57,53.6c0,27.5-21.5,53.6-56.8,53.6h-41.8V79.1z M124.3,172.4h24.5c34.9,0,42.9-26.5,42.9-39.7c0-21.5-13.7-39.7-43.7-39.7h-23.7V172.4z}
                 svg{M88.7,56.8c0,5.5-4.5,10.1-10.1,10.1c-5.6,0-10.1-4.6-10.1-10.1c0-5.6,4.5-10.1,10.1-10.1C84.2,46.7,88.7,51.3,88.7,56.8z};
  }
}

\newcommand\orcidicon[1]{\href{https://orcid.org/#1}{\mbox{\scalerel*{
\begin{tikzpicture}[yscale=-1,transform shape]
\pic{orcidlogo};
\end{tikzpicture}
}{|}}}}

\makeatletter
\def\blfootnote{\xdef\@thefnmark{}\@footnotetext}
\makeatother

\title{\Huge Linking Macro, Meso, and Micro Scales in Multiphase AGN Feeding and Feedback}

\author{Massimo Gaspari$^{1,*,\dagger,\small\orcidicon{0000-0003-2754-9258}}$, Francesco Tombesi$^{2,3,\small\orcidicon{0000-0002-6562-8654}}$, Massimo Cappi$^{4,\small\orcidicon{0000-0001-6966-8920}}$}

\usepackage[breaklinks,colorlinks,citecolor=blue]{hyperref}

\begin{document}
\maketitle

\blfootnote{
\begin{affiliations}
\hspace{-0.4cm}
Affiliations: 
\item Dept.~of Astrophysical Sciences, Princeton University, 4 Ivy Lane, Princeton, NJ 08544, USA.
\item Dept.~of Physics, University of Rome `Tor Vergata', via della Ricerca Scientifica 1, 00133, Rome, Italy.
\item Dept.~of Astronomy, University of Maryland, College Park, MD, 20742, USA.
\item INAF, Osservatorio di Astrofisica e Scienza dello Spazio, via Piero Gobetti 93/3, 40129 Bologna, Italy. \\
\quad$^*$\,Corresponding author -- e-mail: mgaspari@astro.princeton.edu \\
$^\dagger$\,\textit{Lyman Spitzer Jr.}~Fellow
\end{affiliations}
}

\vspace{-0.385cm}

\begin{abstract}
Supermassive black hole (SMBH) \textit{feeding} and \textit{feedback} processes are often considered as disjoint and studied independently at different scales, both in observations and simulations. 
We encourage to adopt and unify three physically-motivated scales for feeding and feedback (micro\,--\,meso\,--\,macro $\sim$ mpc\,--\,kpc\,--\,Mpc), linking them in a tight multiphase self-regulated loop (Fig.\,\ref{fig:all}). 
We pinpoint the key open questions related to this global SMBH unification problem, while advocating for the extension of novel mechanisms best observed in massive halos (such as chaotic cold accretion) down to low-mass systems. 
To solve such challenges, we provide a set of recommendations that promote a multiscale, multiwavelength, and interdisciplinary community.
\end{abstract}

\vspace{+0.4cm}
In the last decade, increasingly strong efforts have been devoted to understand the role of active galactic nuclei (AGN) during the cosmic evolution of galaxies, groups and clusters of galaxies.
The X-ray observatories (\Chandra, \Newton, \Suzaku, \NuSTAR, \Hitomi) have unveiled spectacular interactions of the central SMBH ($\sim$\,$10^7$\,-\,$10^{10}\,\msun$) with its host environment in the form of X-ray cavities, shocks, metal uplift, and turbulence\supercite{McNamara:2012,Voit:2017}. 
At the same time, the diffuse hot halos of cosmic structures locally condense into multiphase filaments and clouds `raining' onto the central SMBH 
(as detected by \HST, \MUSE, \ALMA, \Magellan, \SOFIA)\supercite{McDonald:2012_Ha,Temi:2018,Tremblay:2018,Combes:2019}.
This rain has been shown to be crucial to grow SMBHs 
via Chaotic Cold Accretion (CCA)\supercite{Gaspari:2013_cca,Prasad:2017,Voit:2018,Gaspari:2019},
recurrently triggering mechanical and/or radiative AGN feedback events.

In the nuclear region, spectroscopical AGN studies have discovered a remarkable diversity of feedback\supercite{Ghisellini:2004,Tombesi:2013,Tombesi:2015,Fiore:2017} and feeding\supercite{Tremblay:2016,Gaspari:2018,Maccagni:2018,Rose:2019,Olivares:2019,Storchi-Bergmann:2019} phenomena, including radio jets, ultrafast outflows (UFOs), warm absorbers, ionized/neutral/molecular outflows, high-velocity infalling CO clouds, and precipitating H$\alpha$ filaments. 
However, AGN feeding detections are still less frequent than feedback features, likely due to observational biases.
With the advent of next-generation telescopes (\JWST, \ELT, \SKA, \XRISM, \Athena)
and massively parallel magneto-hydrodynamic (MHD) simulations, we will probe multiphase inflows and outflows in cosmic structures of remarkably different masses, morphologies, and ages. Linking the macro to micro scales of feeding and feedback is thus vital to understand the (co)evolution\supercite{Gaspari:2017_uni} of SMBHs and host structures.

\begin{figure*}
\hspace{-0.2cm}
\subfigure{\includegraphics[width=1.05\textwidth]{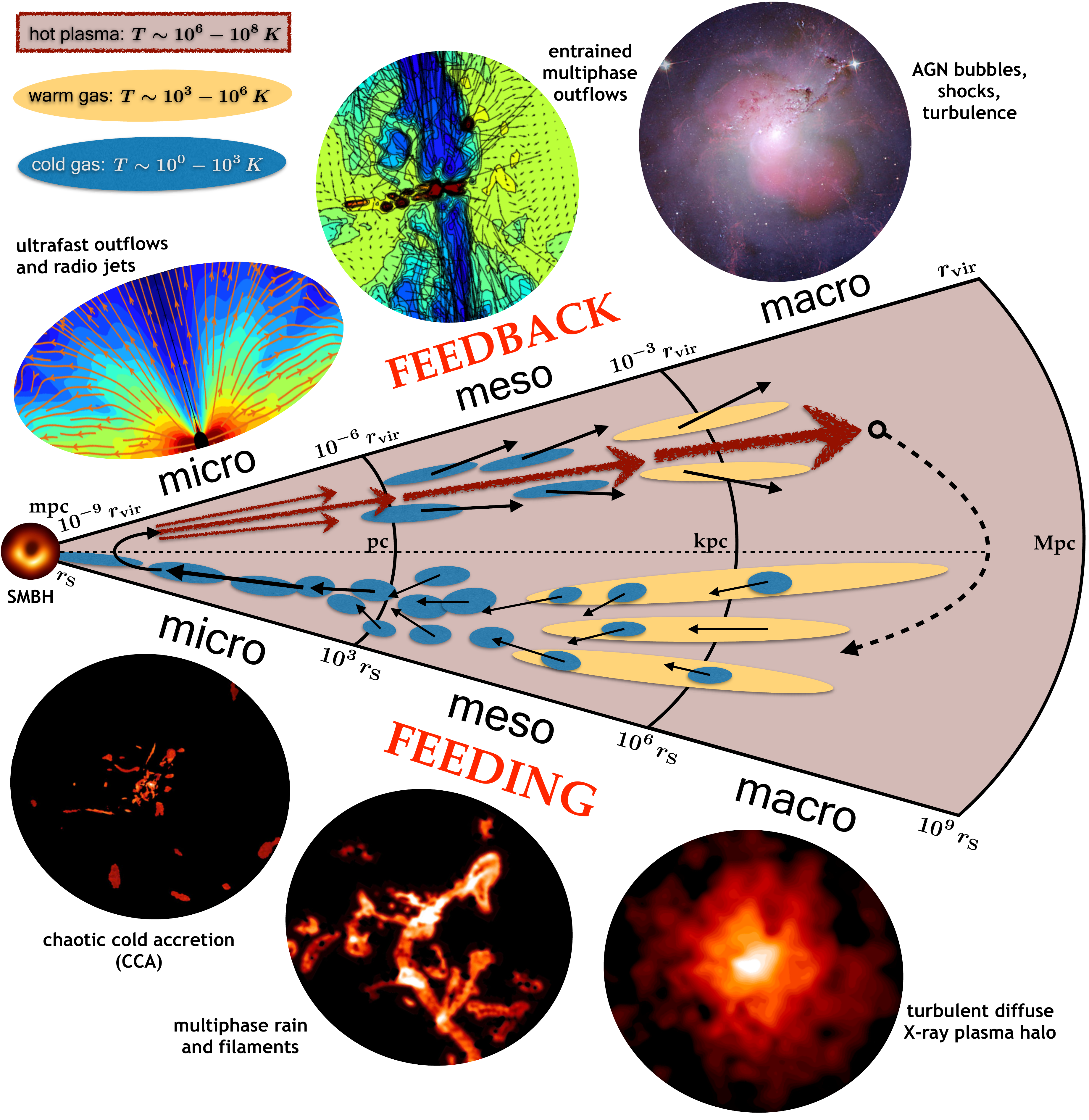}}
\caption{\textbf{The self-regulated multiphase AGN feeding and feedback cycle}. The diagram highlights the three key unification scales (micro -- meso -- macro), which cover a geometric increase of roughly three orders of magnitude each. The macro halo is either a galaxy, group, or cluster and the normalization length is either its virial or Schwarzschild radius (the latter has been directly imaged by the EHT\supercite{EHT:2019I} -- see the adaptation in the middle left inset).
The lower insets show crucial phases of the feeding cycle, in particular the multiphase condensation rain 
out of the turbulent X-ray plasma halo and consequent CCA phase growing the central SMBH (adapted from ref\supercite{Gaspari:2017_cca}).
The upper insets capture key phases of the feedback cycle, i.e., the generation of hot X-ray UFOs and collimated relativistic jets, the entrainment of multiphase ambient gas (or in-situ formation; adapted from refs\supercite{Gaspari:2013_rev,Gaspari:2017_uni}), and the final AGN heating deposition via bubbles, shocks, and turbulence (Perseus image credit: ESA/Hubble Media).  
The multiphase feeding and feedback processes loop for hundreds of cycles during the whole Hubble time.  
}
\label{fig:all}       
\end{figure*}

Studies of SMBH feeding and feedback processes are often disjoint, in terms of approaches, communities, scales, and  wavebands. 
We thus advocate for the joint investigation of both processes in simulations and observations.
For coherence, we suggest to adopt and link three major scales (`micro', `meso', `macro' -- see Fig.\,\ref{fig:all}) defined relatively to the Schwarzschild radius:
\begin{equation} \label{rs}
\rs \equiv \frac{2\,G\mbh}{c^2} \simeq (1\,{\rm mpc})\,\mbhn \simeq (1\,{\rm mpc})\,\txk^2,
\end{equation}
where 
$\mbhn \equiv \mbh/10^{10}\,\msun$ is the SMBH mass and
$\txk\equiv T_{\rm x}/2.5\,{\rm keV}$ is the X-ray plasma halo temperature
(${\rm 1\,mpc = 10^{-3}\,pc \simeq 3\times10^{15}\,cm}$). 
A novel observational finding\supercite{Gaspari:2019} shows that SMBHs are most tightly linked to the properties of the macro X-ray plasma halos (better than the optical/stellar counterparts), $\mbhn \approx \txk^2$, hence the last step in Eq.~\ref{rs}.
We note that $10^{10}\,\msun$ is just a convenient normalization, and does not indicate median population values (which are lower). This scaling applies to diverse environments (central or isolated galaxies, early- or late-type galaxies)\supercite{Gaspari:2019} and can be equally used for SMBHs down to $\sim$\,$10^{7}\,\msun$.
Alternatively, the three scales can be defined as a function of the virial radius
of the group/cluster halo\supercite{Sun:2009a}{}:
\begin{equation} \label{rv}
\rv \equiv \left(\frac{M_{\rm tot, vir}}{4/3\pi\,100 \rho_{\rm c}}\right)^{1/3} \simeq (1.5\,{\rm Mpc})\,\txk^{1/2},
\end{equation}
where $\rho_{\rm c}=3 H^2/(8\pi G)$ is the critical cosmic density of the universe
and $M_{\rm tot, vir}$ is the total virial mass.
The dynamical range between the SMBH and hot-halo scale is 9 dex for a typical group hot halo, $\rv/\rs \simeq 1.5\times10^{9}\;\txk^{-3/2}$, which we use below as simple guide (the range can stretch by another 2 dex down to isolated galaxies).

We suggest to tackle three geometrically increasing \textit{radial} sub-regimes and then reconstruct the full problem (Fig.\,\ref{fig:all}).
Below, we collect several open questions (along with useful insights) that are key to reach the \textit{unification} of SMBH feeding and feedback.
Given the higher photon count, more significant detections arise from massive halos/SMBHs, thus we inevitably give more weight on related mechanisms (such as CCA and precipitation); however, this Comment advocates for their extension and investigation in lower-mass/poorer systems, as any galaxy is expected to grow a gaseous reservoir.

\vspace{+0.5cm}

\noindent
1. \textbf{MICRO}:
$[\,10^0-10^3\,\rs$; $10^{-9}-10^{-6}\,\rv$; \textbf{mpc\,--\,pc}$\,]$
\begin{itemize}
\item What is the main driver of accretion onto the SMBH horizon ($\dot \mbh$): cold or hot mode, chaotic or smooth feeding?  
What is the interplay between classic models (thin/slim disks, radiatively-inefficient accretion flows)\supercite{Shakura:1973,Narayan:1994} and newly discovered mechanisms? E.g.,
CCA can recurrently boost over $100\times$ the classic accretion rates\supercite{Gaspari:2017_cca}. 
Do they vary with the SMBH mass?

\item Is the ubiquitously observed optical/UV/X-ray AGN variability 
related to $\dot \mbh$ or disk instabilities (thermal or magnetorotational)? 
E.g., the CCA flickering rain induces large self-similar variability (power spectral density $\propto \nu^{-1}$) and can alternate the soft/hard (quiescent/quasar) AGN state. 

\item What is the geometry (thin vs.~thick) of the accretion and ejection flows near the SMBH horizon\supercite{Ohsuga:2009}?
Is the inner X-ray AGN corona connected to shocks/magnetic reconnection\supercite{Ghisellini:2004} or an extension of the hot halo? 
What is the effective viscosity $\alpha$ of the micro accretion disk/flow? 

\item What is the main SMBH ejection mechanism?
Magnetic towers\supercite{Fukumura:2010} and X-ray/UV radiation pressure\supercite{Proga:2000} appear to be key candidates at low and high Eddington ratios, respectively
(where $\dot M_{\rm Edd} \equiv L_{\rm Edd}/(0.1\,c^2)\simeq (23\,\myr)\,M_{\bullet,9}$).

\item How is the kinetic AGN power ($\dot E_{\rm k}$) partitioned into wide subrelativistic ($v_{\rm out}$\,$\sim$\,$10^4\,\kms$) massive (X-ray) UFOs and collimated relativistic light (radio) jets?
What is the distribution of the related horizon mechanical efficiency $\varepsilon_{\rm m}= \dot E_{\rm k}/(\dot \mbh c^2)$?

\item How are the micro $\dot M_\bullet$ and $\dot M_{\rm out}$ linked? 
Given the quasi ubiquity of either AGN outflows or jets, only a small fraction of gas is likely sinked through the SMBH horizon (e.g., over 90\% of the inflowing mass is expected to be re-ejected during CCA)\supercite{Gaspari:2017_uni}.

\item Will the new multiwavelength observatories detect micro `ultrafast inflows' (UFI; $v_{\rm in} \sim 10^3\,\kms$) -- the counterparts of UFOs -- and balance the detections of feeding and feedback features?\\ 

\end{itemize}

\noindent
2. \textbf{MESO}:    
$[\,10^3-10^6\,\rs$; $10^{-6}-10^{-3}\,\rv$; \textbf{pc\,--\,kpc}$\,]$
\begin{itemize}

\item Which values of the AGN kinetic energy rate ($\dot E_{\rm k}= {1/2}\,\dot M_{\rm out} v_{\rm out}^2$) and momentum rate ($\dot p=\dot M_{\rm out} v_{\rm out}$) are required to drive sufficient AGN feedback? A key requirement seems that $\dot E_{\rm k}$ must balance the hot-halo $L_{\rm x}$, 
in order to avoid a cooling flow catastrophe.\supercite{Gaspari:2017_uni}

\item How are different phases of the meso AGN-outflow phenomenon connected? 
Energy conservation seems to be key to shape co-spatial multiphase outflows\supercite{Tombesi:2015} ($v_{\rm out} \propto \dot M_{\rm out}^{-1/2}$), leading to slower ($v_{\rm out}\sim 5\times10^3 - 10^2\;\kms$) and more massive ($\dot M_{\rm out}\sim 1-10^3\;\myr$) meso outflows, from the ionized (X-ray/UV), to neutral (IR/21cm) and molecular (radio) phase.
What is the incidence of purely momentum-conserving outflows ($v_{\rm out} \propto  \dot M_{\rm out}^{-1}$)? 

\item What are the probability distributions of the mass loading rates ($\dot M_{\rm out}$), velocities ($v_{\rm out}$), and ionization parameters over large unbiased samples?
We are lacking a complete sample of X-ray/UV/optical multiphase outflows over a large $\dot M_{\rm Edd}$ and redshift space.

\item What is the mass exchange between the chaotic rain and the molecular torus/accretion disk?
What is the role of bars/spirals and gravitational torques\supercite{Storchi-Bergmann:2019}?
The classic separation between coherent structures and clumpy inflows/outflows is weakening; 
e.g., turbulence can intertwine circulating and outflowing gas, while  
broadening the spectral lines. 

\item What is the interplay of fundamental physics in shaping the multiphase meso inflows/outflows? 
The combination of nonlinear physical processes (e.g., cooling, heating, turbulence) may drive new, unexpected 
accretion/ejection mechanisms. 

\item Can we find new direct observational probes of CCA (such as the recent ALMA obscuration `shadows'\supercite{Rose:2019})?
Are obscured AGN linked to precipitation? What is the role of dusty clouds and CO disks during CCA?

\item How does the molecular gas form in hot outflows? 
It is unclear whether fast outflows can directly entrain the ambient medium or whether the cold gas condenses in-situ via thermal or MHD instabilities.
\\
\end{itemize}

\noindent
3. \textbf{MACRO}: 
$[\,10^6-10^9\,\rs$; $10^{-3}-10^{0}\,\rv$; \textbf{kpc\,--\,Mpc}$\,]$
\begin{itemize}
\item How is the feedback energy deposited within the circumgalactic (CGM), intragroup (IGrM), or intracluster (ICM) medium? While X-ray data and simulations suggest that macro AGN feedback acts via buoyant cavities, shocks, and turbulence ($\sigma_v \sim 200\,\kms$)\supercite{McNamara:2012,Gaspari:2013_rev}, the detailed energy transfer and composition (e.g., cosmic rays) are still unclear. 

\item What is the effect of plasma kinetics on the diffuse X-ray/UV macro halo? Plasma instabilities (firehose, mirrors) may play a role in shaping the final AGN feedback thermalization, e.g., by altering the (anisotropic) viscosity and conductivity. 

\item What is the origin and long-term evolution of the multiphase filaments extending out to 100 kpc?
The tight spatial and kinematical (e.g., ensemble line broadening) correlations found between optical/IR (\HST, \MUSE, \textit{VIMOS}, \Magellan) and radio (\ALMA, \textit{WSRT}, \textit{VLA}) data suggest that the H$\alpha$ filaments and CO clouds originate from the halo rain\supercite{McDonald:2012_Ha,Tremblay:2018,Combes:2019}.

\item Do galaxy/SMBH mergers (via the hierarchical $\Lambda$CDM assembly\supercite{Hopkins:2010}) and ram-pressure stripping significantly affect the evolution of macro AGN feeding and feedback processes?

\item What is the multiphase AGN feedback duty cycle over the entire Hubble time? 
How does it correlate with the (quenched) star formation rates in different wavelengths (UV, optical, IR)?
Does the CCA flickering shape the duty cycle at $z > 1$?

\item Can we find minimal {\it physical} parameters to capture AGN feedback in macro simulations? Current cosmological simulations rely on many ad-hoc numerical parameters, which dramatically reduce their predictive power.

\item Does AGN feedback operate in all galaxies?
While X-ray telescopes detect ubiquitous AGN feedback imprints in the gas-rich ICM/IGrM of massive/central galaxies, evidences in spiral/isolated galaxies and dwarfs are still difficult to acquire (due to the low photon count). 
Does the CCA-driven self-regulation scale down to the poor systems\supercite{Maccagni:2018,Gaspari:2018}, given that the related fundamental physics is expected to operate in all environments?

\end{itemize}
\vspace{+0.5cm}

In order to ultimately solve the above key scientific questions,
we advocate for the vital integration between theory and observations. 
While data acquisition and reduction have become increasingly more complex, requiring dedicated technical teams and billion-dollar observatories, numerical studies necessitate massive high-performance computing (HPC) resources and expertise.
The consequent narrower focuses, unbalanced funding allocations, and discrepant timelines
are creating a growing disconnect between the two communities. However, without theoretical predictions, observations become a mere collection of data; without observations, theory can drift into dream land. 
We  thus propose to consider the following recommendations to reach the above long-term objectives:

\begin{itemize}
\item
Include \textit{both} the detailed simulations/analytics and observational tests when performing (multiscale) astrophysical investigations.
Generating accurate synthetic observations with end-to-end pipelines\supercite{Roncarelli:2018} is crucial to compare predictions and data with the same degree of uncertainties and biases, including instrumental/background noise, projection effects, band filters, resolution, field of view, and exposure.
An example is the ongoing \textit{BlackHoleWeather}\supercite{Gaspari:2019} program aimed to tackle the above AGN feeding and feedback problems with this complementary methodology.
At the same time, extracting detailed observables from HPC simulations is key to guide the development of the next-generation instrumentation. 

\item
Encourage committees (e.g., ERC, NSF, NASA, ESA) to grant financial support for programs that address multiscale approaches and leverage interdisciplinary expertise, in particular aimed at linking the above micro to macro scales. 
A more balanced partition of funding between theoretical and observational studies is key to be achieved, the latter currently exceeding the former. 

\item
Encourage peer-review panels to better appreciate and approve multiwavelength proposals (e.g., concurrently using X-ray, optical, and radio telescopes), as well as archival studies supported by numerical simulations. 

\item
Besides institutions and groups, individual researchers can substantially impact the integration of different expertise, in particular by leading the organization of recurrent, highly collaborative workshops (such as those hosted by the SCfA) aimed at bringing together different fields (e.g.,~{\small \url{https://www.sexten-cfa.eu/event/multiphase-agn-feeding-feedback}}). 

\end{itemize}

In conclusion, the concrete and committed adoption of a multiscale simulation--observation integration, resource allocation balancing, and interdisciplinary collaboration initiatives will enable us, as a community, to shed light on the above open problems in the upcoming two decades and to fully leverage the related groundbreaking multi-messenger missions (\Athena, \XRISM, \JWST, \ELT, \SKA, \EHT, \LSST, \LISA), ultimately achieving a unified theory of multiphase AGN feeding and feedback (Fig.\,\ref{fig:all}). 

{
\small 
\vspace{+0.5cm}
\noindent
{\bf Acknowledgments.} 
M.\,G.\ is supported by the \textit{Lyman Spitzer Jr.}~Fellowship (Princeton University) and by NASA Chandra GO7-18121X/GO8-19104X/GO9-20114X and HST GO-15890.020-A grants. 
F.\,T.\ acknowledges support by the `Programma per Giovani Ricercatori -- Rita Levi Montalcini' (2014). 
We are very grateful to all the participants of the `Multiphase AGN Feeding \& Feedback' workshop 
(held at the Sexten Center for Astrophysics -- SCfA -- in Sesto, Italy; {\small \url{https://www.sexten-cfa.eu/event/multiphase-agn-feeding-feedback}})
for stimulating presentations that inspired part of this work, and look forward to its forthcoming second chapter 
({\small \url{https://www.sexten-cfa.eu/event/multiphase-agn-feeding-feedback-ii-linking-the-micro-to-macro-scales-in-galaxies-groups-and-clusters}}).

\printbibliography
}

\end{document}